\documentclass[12pt]{article}
\usepackage{graphicx}
\begin{document}

\centerline{\large \bf 
Why are diploid genomes widespread and dominant mutations rare?}

\bigskip
\centerline{Diana Garncarz$^1$, Stanislaw Cebrat$^1$, Dietrich Stauffer$^{1,2}$ 
and Klaus Blindert$^2$}

\bigskip

\noindent
$^1$Department of Genomics, Institute of Genetics and Microbiology, 
University of Wroc{\l}aw, 

\noindent
ul. Przybyszewskiego 63/77, PL-54148 Wroc{\l}aw, Poland\\

\noindent
$^2$ Institute for Theoretical Physics, Cologne University, D-50923 K\"oln,
Euroland.  

\bigskip
\bigskip
Abstract

\noindent
We have used the sexual Penna ageing model to show that the relation between
dominance and recessiveness could be a force which optimizes the genome
size. While the possibility of complementation of the damaged allele by its
functional counterparts (recessiveness) leads to the redundancy of genetic
information, the dominant effect of defective genes tends to diminish the
number of alleles fulfilling the same function. Playing with the fraction of
dominant loci in the genome it is possible to obtain the condition where the
diploid state of the genome is optimal.
If the status of each bit position as dominant or recessive 
mutations is changed for each individual randomly and rarely, then after a long 
time a stationary equilibrium of many recessive and few dominant loci is 
established in the sexual Penna model. This effect vanishes if the same 
changing distribution of dominant loci applies to all individuals.

\section{Introduction}

Recent advances in the genome analyses have indicated that genomes of the
closely related species can differ substantially in their sizes. To stress
the flexibility of the genome structure and size, the phenomenon is called
sometimes "the DNA (genome) in flux". There are many mechanisms, at
different levels of genome organization, which influence the genome size.
Some very sophisticated mechanisms rearrange specific sites inside the
genome during the development. These mechanisms eliminate some sequences
from the genome but the changes are not inherited because the programmed
excisions take place in the somatic cells being out of the germ line, like
rearrangements of genes coding for immunoglobulins or T cell receptors in
lymphocytes \cite{Tonegawa}. Some DNA excisions lead to generation of the
new functions of the cell but simultaneously stop the cell divisions like in
the case of generation of the nitrogenase gene in blue algae \cite{Golden}.
Some changes could be very minute, adding or eliminating single nucleotides,
or they could be very substantial, adding or eliminating genes, clusters of
genes, whole haplotypes or even duplicating whole genomes. Additions or
deletions of small numbers of nucleotides, if inside coding sequences,
usually are deleterious for the functions of the sequences (genes) and the
negative selection eliminates the mutants. Addition of a complete coding
sequence, called gene duplication, produces redundant information in the
genome (paralog genes). The additional copy of the gene can stay in the
genome if it complements the function of its homolog, enhances this function
or by any other means helps the host to compete with other organisms. If the
copy is dispensable, it is lost during later evolution due to neutral
selection.

\begin{figure}[hbt]
\begin{center}
\includegraphics[scale=0.8]{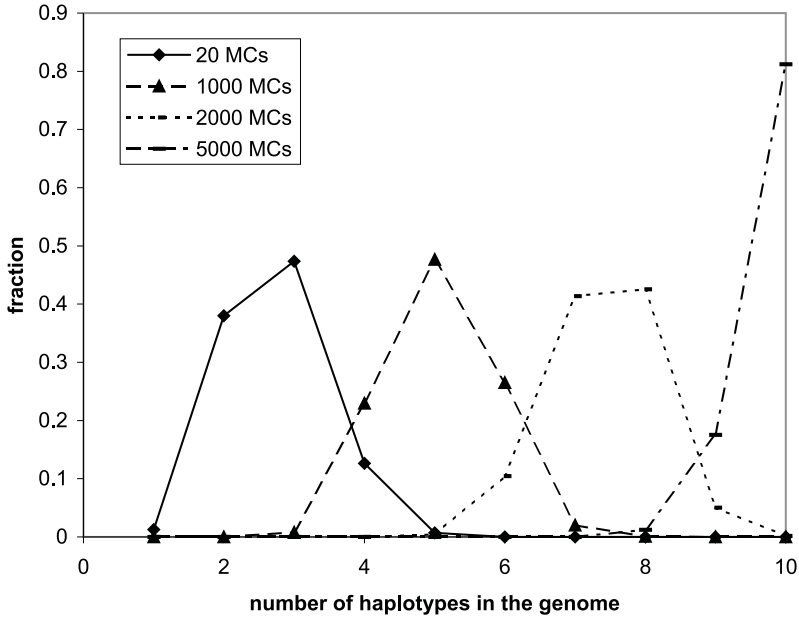}
\end{center}
\caption{
Diagrams showing the increasing numbers of haplotypes in the genomes
during the simulations when all defective genes were recessive.
}
\end{figure}

\begin{figure}[hbt]
\begin{center}
\includegraphics[scale=0.8]{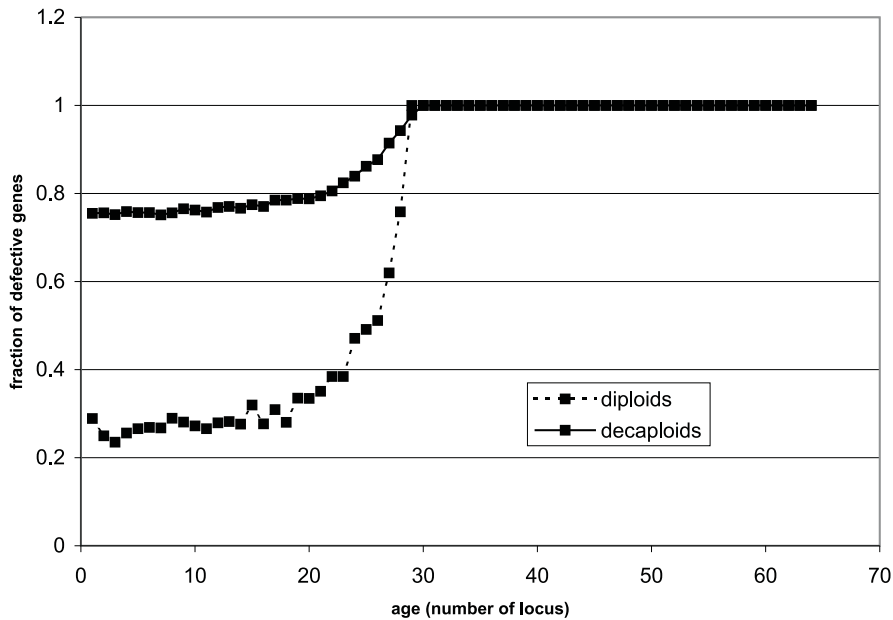}
\end{center}
\caption{
Distribution of defective genes in the genetic pool of the
population depending on time of their expression for diploid and decaploid
genomes (two or ten haplotypes in the genome, respectively).
}
\end{figure}

\begin{figure}[hbt]
\begin{center}
\includegraphics[scale=0.8]{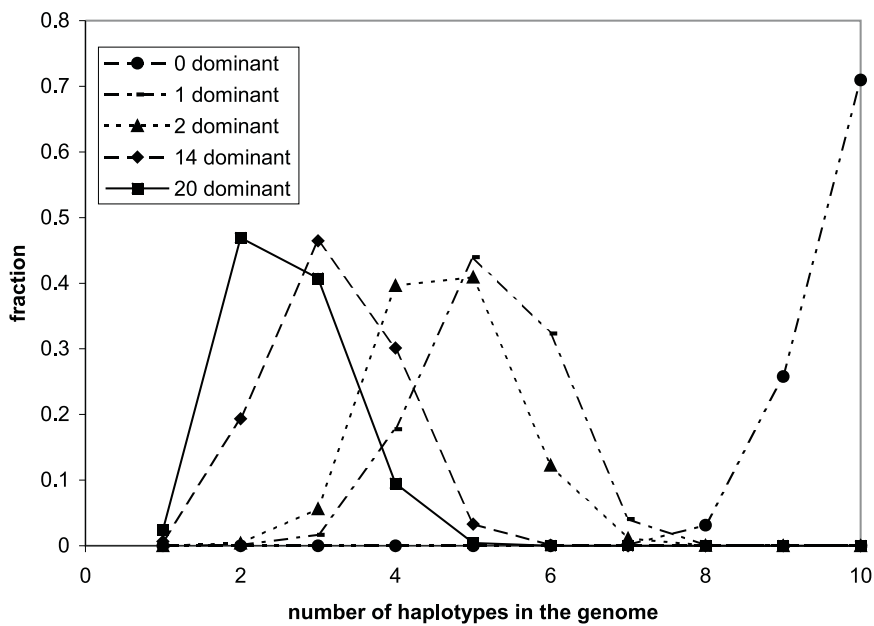}
\end{center}
\caption{
Diagrams showing the numbers of haplotypes in the individual genomes
with different number of declared dominant loci (dominant loci were situated
as the first ones in the genomes).
}
\end{figure}

\begin{figure}[hbt]
\begin{center}
\includegraphics[scale=0.8]{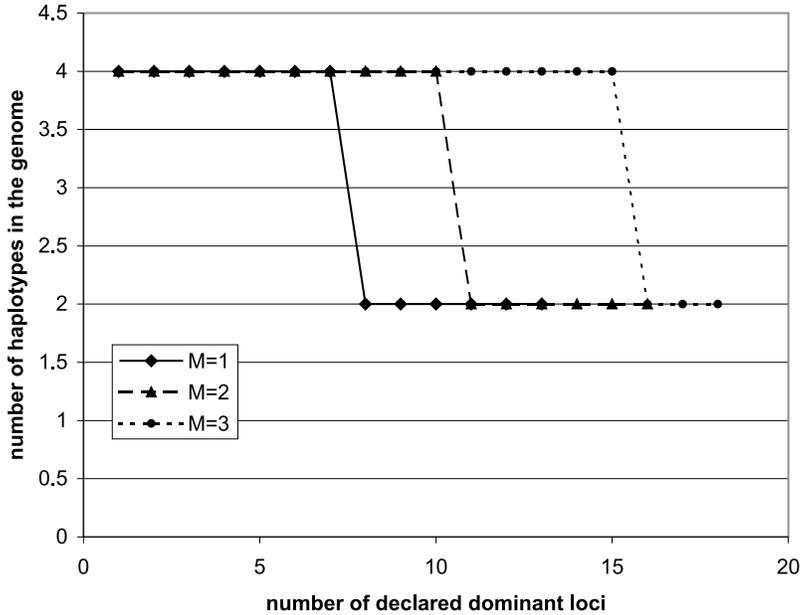}
\end{center}
\caption{
Transition from tetraploid state to diploid state when the number of
declared  dominant loci increased for different mutational pressure.
}
\end{figure}

\begin{figure}[hbt]
\begin{center}
\includegraphics[scale=0.8]{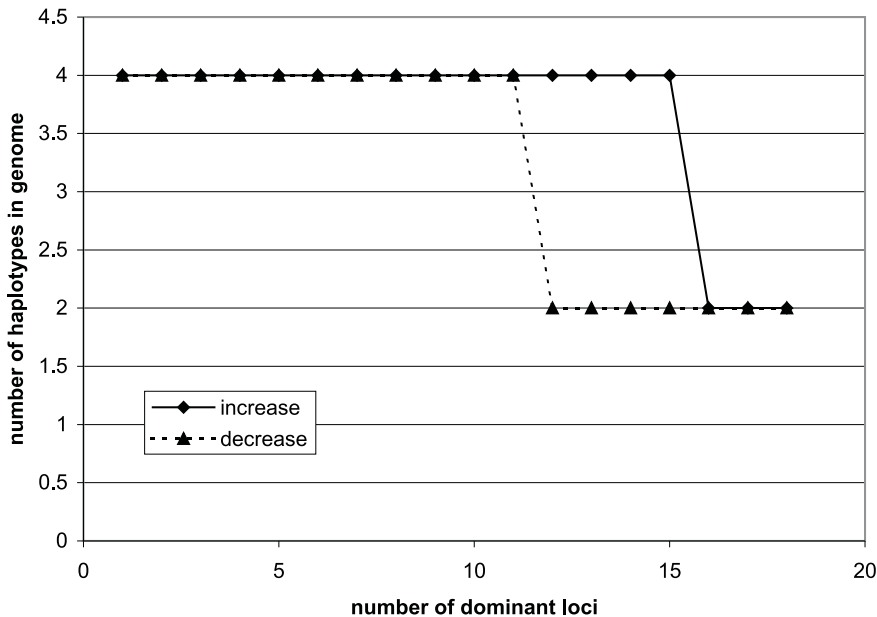}
\end{center}
\caption{
Stabilization of diploid state. After increase in the number of
dominant loci to 16 all genomes became diploid, but they stayed diploid when
the number of dominant loci was decreased in these genomes to 12. Mutation
rate in these simulations: $M=1$.
}
\end{figure}

\begin{figure}[hbt]
\begin{center}
\includegraphics[scale=0.9]{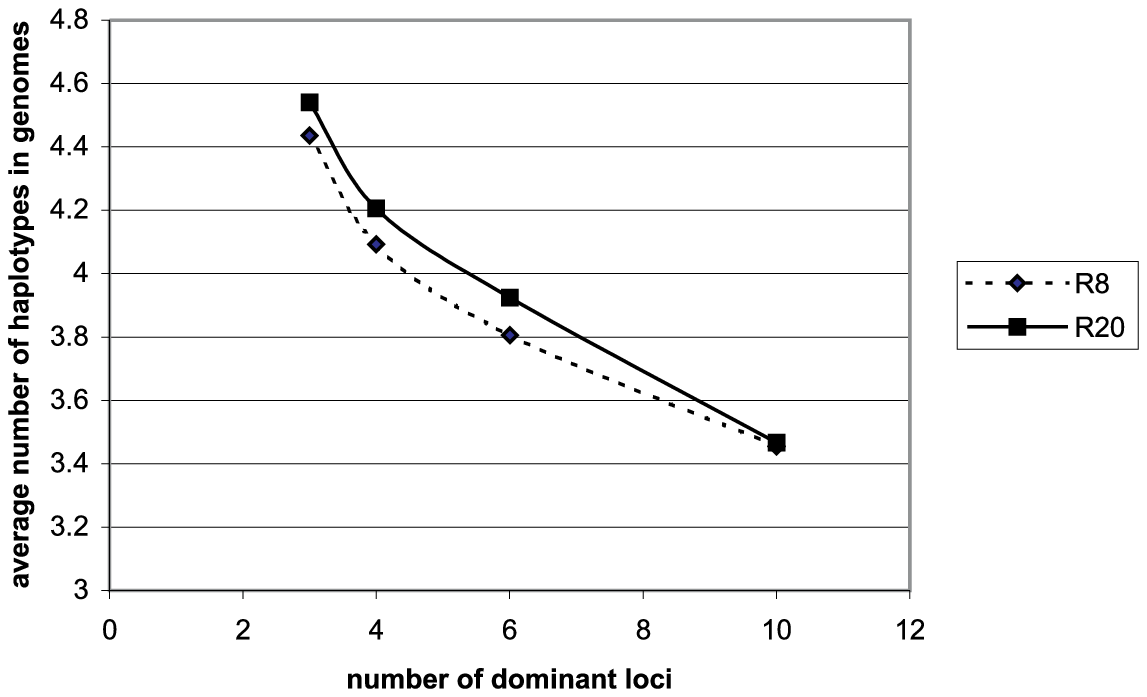}
\end{center}
\caption{
Relation between the number of dominant loci and the average ploidy
of genomes for different minimum reproduction ages.
}
\end{figure}

In this paper we are going to discuss and to model only some major events
which duplicate the whole genomic information. A single, complete set of
genomic information is called a haplotype. Organisms or cells possessing one
such set are called haploids. If an organism possesses two homologous
haplotypes, it is called diploid, if more, polyploid. In some instances
whole genomes can be duplicated. Such a duplication of a haploid genome is
called diploidization. In fact some normal, developmental processes like
sexual reproduction of yeasts could be considered as diploidization. Two
haploid cells with homologous haplotypes fuse together and form one diploid
cell. In this case the process could be considered reversible - diploid
cells even after many generations can again produce haploid cells. If we
neglect the existence of homologous genes which could fulfill the same
function in one haplotype (i.e. duplicated genes, paralogs) we can assume
that each function of the haploid organism is performed by a single copy of
a gene. If the gene is destroyed, its function is also destroyed and if it
is important for survival, the whole organism is eliminated. In diploid
organisms each gene exists in two copies. Destroying one copy of the gene
does not necessarily lead to depriving the organism of the function. The
other copy of the gene can "complement" the function of the homologous
destroyed gene. This phenomenon is called recessiveness. Thus, it seems
obvious that diploid organisms should be much more robust and resistant to
the mutational pressure than haploids because both homologous genes
(localized at the corresponding positions in the haploid genomes: loci)
have to be destroyed to eliminate the function. The recessiveness is not the
only advantage which can be provided by the diploidization. The homologous
genes in one locus (alleles) could be slightly different and could provide
slightly different genetic information which could be also advantageous for
the organism, as in the case of many loci involved in the immunological
processes. If we assume that diploidization is such an ingenious invention
of the Nature, why is the higher polyploidization not exploited more often?
In fact some higher polyploidizations have been observed many times,
especially in plant genome evolution \cite{Soltis93}, \cite{Soltis99}.
The polyploidization events usually are followed by the genome reduction
which leads back to the diploidization \cite{Wolfe} and the only effect of
the ancient polyploidization is the relatively high number of paralogs, some
of them can lose their functions (nonfunctionalization) \cite{Wendel} or some
of them can preserve their function for long time \cite{Lynch}.

Computer modeling of polyploidization has shown that multiplying the whole
set of haplotypes could be profitable for organisms \cite{Alle}. If the
recessiveness of all mutations was assumed and the polyploidy was introduced
as an evolving feature of the genomes, the number of haplotypes in organisms
had a tendency to grow to the infinity. Nevertheless, parallel to the
polyploidization, the genetic load (the fraction of defective alleles) in
these genomes also grew. One can conclude that such a polyploidization leads
to the accumulation of a high fraction of redundant information which is
allowed to be destroyed by mutations. Unfortunately, in Nature, keeping even
destroyed information is costly. High energetic costs are paid for the DNA
replication and often for the expression of wrong or dispensable
information. When the costs of genome replication were introduced into the
model by elongating the generation time of the organism proportionally to
its genome, the tendency to polyploidization was reduced \cite{Alle}. On
the other hand, declaring all loci recessive is an oversimplification. In
fact, not all mutations are recessive and sometimes the function of a single
defective copy of a gene cannot be complemented by its wild allele. There
are known mutations in the single copy of a gene which lead to deleterious,
even lethal effects (such mutations are called dominant). Among the numerous
examples of dominant mutations is a dynamic mutation in the gene responsible
for Huntington chorea or mutations in the cellular protooncogenes. Mutations
in the later group of genes, even in somatic cells, could induce cancer
followed by death of the whole multicellular organism. Thus, it seems that
polyploidization could be restricted also by the effect of dominance. Since
the probability of mutation in a single locus linearly depends on the number
of alleles in the locus, the deleterious effects of dominant mutations
should be enhanced in the polyploid genomes. Sousa et al. \cite{Sousa} have
shown such an effect in the computer simulations when analyzing different
strategies of sexual reproduction. They introduced the dominant loci into
the simulated genomes and showed that a triploid phase is
not an efficient strategy in reproduction.

To simulate the effect of dominance in the phenomenon of the genome
polyploidization we have used a modification of the sexual Penna ageing model
\cite{Penna} which enables the study of the influence of different genetic
parameters on the population size, age distribution or structure of its
genetic pool.

\section{Model}
In our version of the Penna model each individual is represented by its
genome which could consist of different number of haplotypes. One haplotype
is a bitstring $64$ bits long. The value of each bit could be $0$, which
represents the wild type (correct) gene or $1$, which represents a defective
gene. The bits in the strings are numbered consecutively and bits placed at
the same position (locus) in different strings represent alleles. If a given
locus is declared a dominant one - even one bit set for $1$ at this
position determines the defective phenotype of the locus. If a locus is
declared a recessive one, it means that all bits at this position have to
be set for $1$ to determine the defective phenotype. Otherwise, the
phenotype of this locus is correct. Like in the standard Penna model, genes
in the loci are switched on chronologically, in the first step all alleles
at the first locus are switched on, in the second step the alleles of the
second locus are switched on and so on. If the declared number $T$ of
defective phenotypes has been expressed, the organism dies. If before
dying, the organism reaches the minimum reproduction age, it produces two
gametes (see below). One of these two gametes randomly drawn is joined
with a gamete produced by another individual of reproductive age and
forms a "newborn" which will be one step old in the next step if it survives
the "Verhulst test". To avoid the overcrowding of the environment, the
logistic equation of Verhulst is introduced to control the birthrate;
$V=1-N_t/N_{\max}$, where $V$ describes the survival probability of the newborn,
$N_t$ corresponds to the actual size of the population and $N_{\max}$ is called 
the maximum capacity of the environment. If a newborn passes the Verhulst test
it could die in the future because of too many defective phenotypes switched
on or because of reaching the maximum age which equals the number of bits in
the string.

Simulations start with a fraction of haploids - 0.25, diploids - 0.25 and
triploids - 0.5, all bits are set for $0$, the threshold $T$ is declared
$3$, the minimum reproduction age is set for $8$ and birth rate is $1$.
In this version of the model the most critical is the production of gametes
and newborns. Before this procedure, a declared number of mutations $M$ is
introduced into randomly chosen loci of each haplotype of the reproducing
parent. If the chosen bit is $0$ it is replaced by $1$ if it is already $1$
it stays $1$ which means that there are no reversions. Next, haplotypes are
randomly paired and one recombination (crossover) takes place at a random
point in each pair. If the number of haplotypes in the individual genome is
not even, one haplotype does not recombine. After recombination haplotypes
from each pair are assorted randomly into two gametes. Note that if the
number of haplotypes in the parental genome was not even, one gamete would
have more haplotypes than the other one. After forming the individual by the
fusion of two gametes, its genome could shrink or grow with equal
probability by random choice of one haplotype and its elimination or
replication.

\section{Results and Discussion}
The first simulations with all loci declared recessive have shown that the
number of haplotypes in the genomes has a tendency to grow to infinity
supporting the results previously obtained by Alle \cite{Alle}. To save 
computer time we have set the upper limit of polyploidy at 10 (Fig. 1).
There are two versions of introducing the limit: in the first one, when the
polyploidy of a newborn after gamete fusion reaches 10 it can loose one
haplotype with the probability 0.5 but it cannot gain another one. In the
second version polyploidy stays at 10. After introducing the first rule,
most of individuals reach "decaploidy" but there is always a fraction of
individuals with less than 10 haplotypes. In the second version, all
individuals in the populations reach decaploidy. This is not the only
difference between the versions. In the standard Penna model simulations, at
equilibrium, the genetic pool of the population is characterized by a
specific gradient of fractions of defective genes. The fractions stay
constant and low for all loci expressed before the minimum reproduction age,
grow in the loci expressed after the minimum reproduction age and reach $1$
for the last loci in the bitstrings. This characteristic structure of the
genetic pool is observed in the populations when the individuals reaching
the upper limit of haplotypes stay with this number and when in equilibrium
all genomes are decaploid (Fig. 2). If the number of haplotypes in the
decaploid newborn genomes may be reduced, the population does not reach the
characteristic distribution of defective genes.

The tendency to the unlimited growth of polyploidy can be reduced by
declaring the dominant loci. In the simulations, different numbers of loci
were declared dominant (always at the beginning of the bitstrings). The
diagrams showing the distributions of the number of haplotypes in the
genomes for populations with different numbers of declared dominant loci are
shown in Fig. 3. The increase in the number of dominant loci is associated
with the decrease in the average number of haplotypes in the genomes.
Since introducing the dominant loci into the genomes reduced the tendency to
polyploidization, in further simulations we have set the upper limit of
polyploidization for $4$. Now, during reproduction, only two types of
gametes could be produced: haploid and diploid. The zygotes could be
diploid, triploid or tetraploid but one haplotype of triploid zygotes was
replicated or lost with equal probability - 0.5. Thus, in the populations
the genomes could be either diploid or tetraploid. When we increased the
declared number of dominant loci we observed transition from tetraploidy to
diploidy (Fig.4).  To reach the pure diploid population in the equilibrium,
a substantial fraction of active loci in the genomes have to be declared
dominant. But when the population already reaches the state of diploidy, it
is stable even when the fraction of dominant loci diminishes. It is also
possible to get the diploid population with a lower fraction of declared
dominant loci when the mutational pressure is increased (Fig. 5).
In the Penna model, the number of active loci in the genome (maximum life
expectancy) grows with increasing minimum reproduction age. We have checked
if the effect of dominance on the evolution of polyploidy depends on the
number of dominant loci or the fraction of dominant loci. To estimate that,
we have compared the average polyploidy in populations with minimum
reproduction age 8 and 20 and the same number of declared dominant loci. The
results suggest that the effect depends on the number of dominant loci
rather than on the fraction of dominant loci (Fig. 6).

\bigskip
\bigskip
\begin{figure}[hbt]
\begin{center}
\includegraphics[angle=-90,scale=0.5]{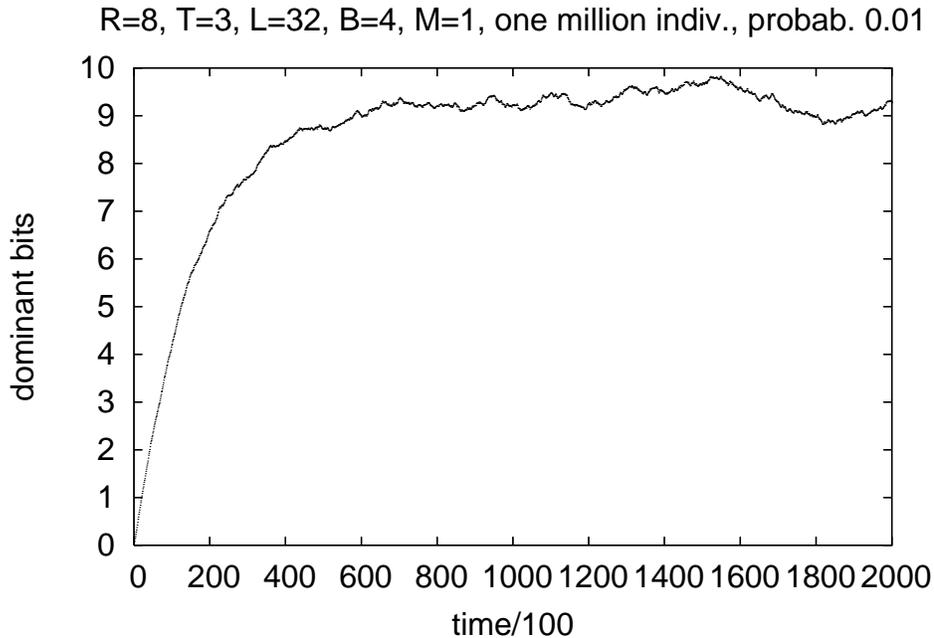}
\end{center}
\caption{Self-organisation of average number $<d>$ of dominant loci, among 
a total of 32 loci, at $p = 0.01$.
}
\end{figure}

\begin{figure}[hbt]
\begin{center}
\includegraphics[angle=-90,scale=0.5]{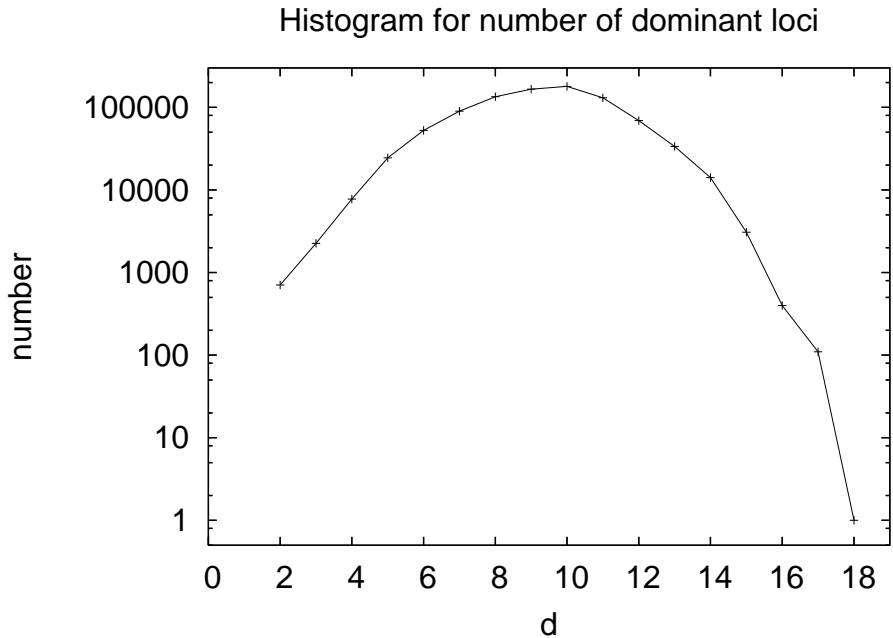}
\end{center}
\caption{Distribution of the number $d$ of dominant loci in same simulation
as in Fig.7.
}
\end{figure}

\begin{figure}[hbt]
\begin{center}
\includegraphics[angle=-90,scale=0.5]{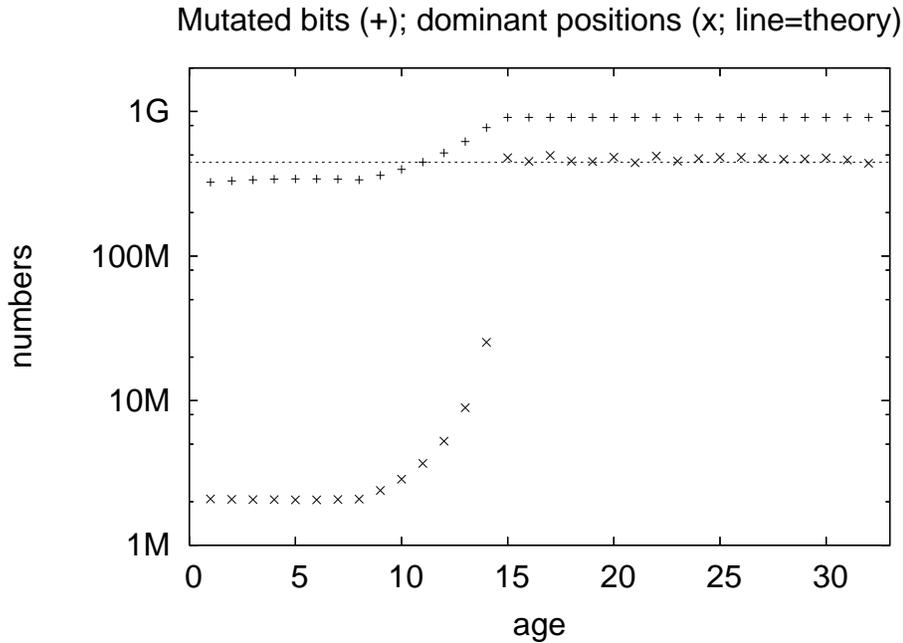}
\end{center}
\caption{Distribution (x) of the number $d$ of dominant loci in same simulation
as in Fig.7. The straight line at 455 million indicates the case of half the 
loci being dominant for this statistics. The plus signs show the number of 
mutated bits: All bits at old age are mutated.
}
\end{figure}

\begin{figure}[hbt]
\begin{center}
\includegraphics[angle=-90,scale=0.5]{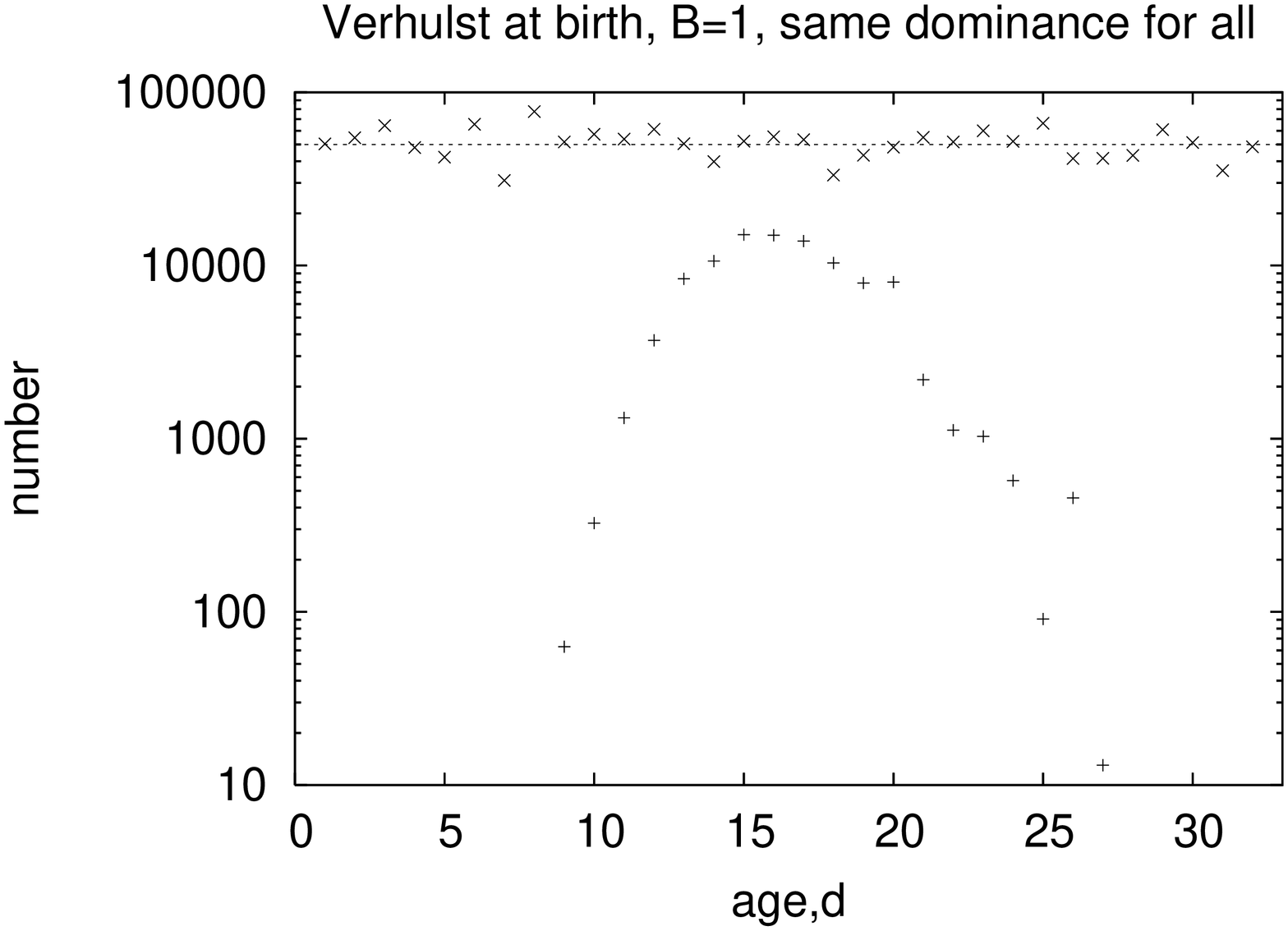}
\end{center}
\caption{Distribution (x) of the number $d$ of dominant loci with Verhulst 
applied to newborns only, $B=1$, and only one distribution of dominant loci for
the whole population. The plus signs show the distribution of the number $d$
of dominant loci in this simulation. The horizontal line correponds to half
the loci dominant and the other half recessive.
}
\end{figure}

\section{Emergence of Dominance}

In the above sections we have shown that an increasing fraction of dominant 
loci forces the organisms to restrict the number of haplotypes 
in the single genome. Now, the question is: what would be the fraction of 
dominant loci if it could freely evolve in the population with 
diploid genomes.

Usually, the characterisation of a locus (bit position) as dominant or recessive
is fixed initially as the same for all individuals: a dominance bit-string has
6 randomly selected bits set to one and the other 26 set to zero if 
bit-strings of length 32 are used. These numbers correspond to the empirical
fact that recessive diseases are much more widespread than dominant ones.

We now undertake a more realistic simulation by letting this distribution of
dominant and recessive mutations self-organize (``emergence'') from the case
where all loci are recessive. Then at each birth, with low probability $p$, a 
maternal locus is selected randomly, independently for each individual, and 
changed for the child
from recessive to dominant or from dominant to recessive. Since this change 
is more complicated than usual mutations, the rate $p$ of change 
should be much smaller than the rate $M=1$ of mutations, taken as one per
iteration and per bit-string. Our other parameters were: minimum age of 
reproduction $R = 8$, lethal threshold for active diseases $T=3$, birth rate
$B=4$ per iteration for all active females, length $L=32$ of bit-strings,
population $10^6$.
 
Fig.7 shows how over many thousand iterations a reasonable average number of
dominant loci emerges from the initial zero. Fig.8 shows the histogram of the
number of dominant loci in each individual; we see a broad and roughly Gaussian
distribution. Fig.9 finally shows, as a function of age (bit position) the 
number of individuals having a dominant locus at that bit position. 

The distribution of dominant loci in the genomes corresponds to selection
pressure exerted on the genes. The first eight loci, expressed before the
minimum reproduction age, under strong selection, are almost free of
dominance (fraction of dominant loci is of the order of mutation rate for
dominance trait). Loci expressed after the minimum reproduction age are
under the gradient of selection pressure which eventually reaches 0 for loci
higher than 15. For these loci which are not under the selection (18 loci
under the set of parameters used for simulations), the average fraction of
dominant loci is 0.5. In fact, Fig. 7 shows that the fraction of
dominant loci in the whole genomes in equilibrium fluctuates between 9 and
near-10, which could be interpreted that self-organization leads to the near-total
avoiding the dominance of deleterious mutations in the genes under selection
pressure.

Biologically it is more realistic to assume the {\it same} distribution of 
dominant and recessive loci for {\it all} individuals, instead of having it
different for each different individual as in Figs.7-9. We now also applied 
the Verhulst deaths due to lack of space and food only to the newborns, and 
no longer as in Figs.7-9 to all ages \cite{verhulst}. As a result, the 
distribution of dominant loci (much worse statistics) is now homogeneous over
all ages, Fig.10. Apparently, selection of the fittest dominance distribution
has become impossible since everybody has the same dominance distribution at
any given time; thus the spread of dominance is not hindered by selection. 
The current prevalence of recessive versus dominant hereditary diseases in
nature does not arise, according to Fig.10, from less dominant loci but from
the death of most carriers of dominant diseases. Indeed, for the fixed 
distribution of 6 dominant loci we found one-bits signalling genetic disease
much rarer at the dominant than at the recessive loci (not shown.) 

The distribution of mutated bits looks similar to the plus signs in Fig.9 (not 
shown). The total population is lower (2.4 million versus 3.4 million) for this
case of time-dependent dominance distribution than for a time-independent 
dominance of six fixed loci, and otherwise identical parameters.

\section{Summary}

Two different but related computer simulation gave another explanation why 
diploid instead of e.g. tetraploid genomes are so widespread for sexual 
reproduction, and how a small but positive fraction of dominant instead of 
recessive mutations can evolve.

This work was done in the frame of European programs COST P10 and GIACS. SC was 
supported by Polish Foundation for Science.

\end{document}